\documentclass[a4paper,preprint,showpacs,amssymb,pre,superscriptaddress]
              {revtex4}

\usepackage{amssymb}
\usepackage{graphicx}

\begin{document}

\title{Van Kampen's expansion approach in an opinion formation
model}

\author{M.S. de la Lama} \email{msanchez@ifca.unican.es}
\affiliation{Instituto de Fisica de Cantabria, Universidad de
Cantabria and CSIC \\ 390005-Santander, Spain}

\author{I.G. Szendro} \email{szendro@ifca.unican.es}
\affiliation{Instituto de Fisica de Cantabria, Universidad de
Cantabria and CSIC \\ 390005-Santander, Spain}

\author{J.R. Iglesias} \email{iglesias@if.ufrgs.br}
\affiliation{Instituto de Fisica, Universidade Federal de Rio Grande
do Sul,\\  91501-970 Porto Alegre, Brazil} \affiliation{Programa de
P\'os Gradua\c{c}\~{a}o em Economia,\\ Universidade Federal de Rio
Grande do Sul, Av. Jo\~{a}o Pessoa 52,\\ 90040-000 Porto Alegre,
Brazil}

\author{H.S. Wio} \email{wio@ifca.unican.es}
\affiliation{Instituto de Fisica de Cantabria, Universidad de
Cantabria and CSIC \\ 390005-Santander, Spain} \affiliation{Centro
At\'omico Bariloche, 8400 San Carlos de Bariloche, Argentina}

%
%
\begin{abstract}
We analyze a simple opinion formation model consisting of two
parties, $A$ and $B$, and a group $I$, of undecided  agents. We
assume that the supporters of parties $A$ and $B$ do not interact
among them, but only interact through the group $I$, and that there
is a nonzero probability of a spontaneous change of opinion ($A
\leftrightarrows I$, $B \leftrightarrows I$). From the master
equation, and via van Kampen's $\Omega$-expansion approach, we have
obtained the ``macroscopic" evolution equation, as well as the
Fokker-Planck equation governing the fluctuations around the
deterministic behavior. Within the same approach, we have also
obtained information about the typical relaxation behavior of small
perturbations.
\end{abstract}

%

\maketitle

\section{Introduction}

The last few years have witnessed a growing interest among
theoretical physicists in complex phenomena in fields departing from
the classical mainstream of physics research. In particular, the
application of statistical physics methods to social phenomena has
been discussed in several reviews
\cite{weidlich1,weidlich2,stauffer00,stauffer00p,Galam000}. Among
these sociological problems, one that has attracted much attention
was the building (or the lack) of consensus. There are many
different models that simulate and analyze the dynamics of such
processes in opinion formation, cultural dynamics, etc
\cite{Galam000,Gal00,complex1,heg01,sznajd01,sznajd02,stauffer1,stauffer2,stauffer3,Castellano,Klemm1,krapi01,Mobilia1,nos,schneider1,JRI}.
Even though in general the models studied in those works are simple
ones, most of the results have been obtained via simulations.
However, it is extremely relevant to have some form of analytical
insight.

In this work we analyze a simple opinion formation model, analogous
to the one studied in \cite{redner3} consisting of two parties, $A$
and $B$, and an ``intermediate"  group $I$, that we call
\textit{undecided  agents}. As in \cite{redner3}, we assume that the
supporters of parties $A$ and $B$ do not interact among them, but
only through their interaction with the group $I$, convincing one of
its members through a Sznajd-like rule similarly to what was
discussed in \cite{sznajd02}, that is within a mean-field treatment.
However, we don't consider that members of $I$ can convince those of
$A$ or $B$, but instead we assume that there is a nonzero
probability of a spontaneous change of opinion from $I$ to the other
two parties and viceversa: $I \leftrightarrows A$, and $\,I
\leftrightarrows B$. We will see that this probability of
spontaneous change of opinion (implying the existence of a
\textit{social temperature} \cite{babinec,weidlich2,last}) inhibits
the possibility of reaching a consensus. Instead of consensus, we
find that each party has some statistical density of supporters, and
there is also a statistical stationary number of undecided  ($I$)
agents.

Our aim is to write a master equation for this toy model, and study
its behavior via van Kampen's $\Omega$-expansion approach
\cite{vKamp}. After determining if, in this case, the conditions for
the validity of using such an approach are fulfilled, and exploiting
it, we could obtain the \textit{macroscopic} evolution equations for
the density of supporters of $A$ and $B$ parties, as well as the
Fokker-Planck equation governing the fluctuations around such
deterministic or macroscopic behavior. The same approach also offers
information about the typical relaxation behavior of small
perturbations around the stationary macroscopic solutions.

The outline of the paper is the following. In the next Section we
present the model, and apply van Kampen's $\Omega$ expansion
approach in order to obtain the \textit{macroscopic} equation and
the Fokker-Planck equation governing the fluctuations around the
macroscopic behavior. In Section 3 we analyze the behavior of the
fluctuations through the study of their mean values and
correlations, and discuss the relaxation time of small
perturbations. In Section 4 we present some typical results and
finally, in Section 5, some general conclusions are summarized.

\section{The model and the approach}

\subsection{Description of the model}

We consider a system composed of three different groups of agents\\
$\triangleright$ supporters of the $A$ party, indicated by $N_A$,\\
$\triangleright$ supporters of the $B$ party, indicated by $N_B$,\\
$\triangleright$ undecided  ones, indicated by $N_I$.\\
The interactions we are going to consider are only between $A$ and
$I$, and $B$ and $I$. That means that we do not include direct
interactions among $A$ and $B$. The different contributions that we
include are\\
$\bullet$ spontaneous transitions $A \to I$, occurring with a rate
$\alpha_1 \, N_A$;\\
$\bullet$ spontaneous transitions $I \to A$, occurring with a rate
$\alpha_2 \, N_I$;\\
$\bullet$ spontaneous transitions $B \to I$, occurring with a rate
$\alpha_3 \, N_B$;\\
$\bullet$ spontaneous transitions $I \to B$, occurring with a rate
$\alpha_4 \, N_I$;\\
$\bullet$ convincing rule $A+I \to 2\,A$, occurring with rate
$\frac{\beta_1}{\Omega} N_A N_I$;\\
$\bullet$ convincing rule $B+I \to 2\,B$, occurring with rate
$\frac{\beta_2}{\Omega} N_B N_I$.\\
As indicated above, here $N_i$ is the number of agents supporting
the party or group ``$i$" (with $i=A,B,I$). We have the constraint
$N_A + N_B + N_I = N$, where $N$ is the total number of agents. Such
a constraint implies that, for fixed $N$, there are only two
independent variables $N_A$ and $N_B$. By using this constraint, the
rates indicated above associated to processes involving $N_I$, could
be written replacing $N_I = (N - N_A - N_B).$

With the above indicated interactions and rates, the master equation
for the probability $P(N_A, N_B,t)$ of having populations $N_A$ and
$N_B$ at time $t$ (due we have had populations $N_A^{o}$ and
$N_B^{o}$ at an initial time $t_o (<t)$), may be written as
\begin{eqnarray}
\frac{\partial}{\partial \,t} P(N_A, N_B,t) = & & \alpha_1 (N_A+1)
P(N_A+1, N_B,t) + \alpha_3 (N_B+1) P(N_A, N_B+1,t) \nonumber \\
&+& \alpha_2 (N-N_A-N_B+1) P(N_A-1, N_B,t) \nonumber \\
&+& \alpha_4 (N-N_A-N_B+1) P(N_A, N_B-1,t) \nonumber \\
&+& \frac{\beta_1}{\Omega}(N_A -1)(N-N_A-N_B+1) P(N_A-1, N_B,t)
\nonumber\\
&+& \frac{\beta_2}{\Omega}(N_B -1) (N-N_A-N_B+1) P(N_A, N_B-1,t)
\nonumber \\
&-& \Bigl[ \alpha_1 N_A + \alpha_3 N_B + \alpha_2 (N-N_A-N_B)
\Bigr. \nonumber \\
& & \,\,\,\,\,\,\,\,\,\,\,\,\,\,\, + \Bigl. \alpha_4 (N-N_A-N_B+1)
\Bigr] P(N_A, N_B,t).
\end{eqnarray}
This is the model master equation to which we will apply van
Kampen's approach \cite{vKamp}.

\subsection{Van Kampen's expansion}

In order to apply van Kampen's approach, as discussed in
\cite{vKamp}, we identify the large parameter $\Omega$ with $N$
(assuming $N \gg 1$); and define the following separation of the
$N_i$'s into a macroscopic part of size $\Omega$, and a
fluctuational part of size $\Omega ^{\frac{1}{2}}$,
\begin{eqnarray}\label{xx1}
N_A & = & \Omega \Psi_A (t) + \Omega^{\frac{1}{2}} \xi_A(t),\nonumber \\
N_B & = & \Omega \Psi_B (t) + \Omega^{\frac{1}{2}} \xi_B(t),
\end{eqnarray}
and define the density $\rho= \frac{N}{\Omega}$ (in our case $\rho =
1$). We also define the ``step operators"
\begin{eqnarray}
\mathbb{E}_i^{1} f(N_i) & = & f(N_i+1), \nonumber \\
\mathbb{E}_i^{-1} f(N_i) & = & f(N_i-1), \nonumber
\end{eqnarray}
with $f(N_i)$ an arbitrary function. Using the forms indicated in
Eqs. (\ref{xx1}), in the limit of $\Omega \gg 1$, the step operators
adopt the differential form \cite{vKamp}
\begin{eqnarray}
\mathbb{E}_i^{\pm 1} = 1 \pm \left( \frac{1}{\Omega}
\right)^{\frac{1}{2}} \frac{\partial}{\partial\, \xi_i} +
\frac{1}{2} \left( \frac{1}{\Omega} \right)
\frac{\partial^{2}}{\partial\, \xi_i^{2}} \pm \ldots,
\end{eqnarray}
with $i=A,B$. Transforming from the old variables $(N_A,N_B)$ to the
new ones $(\xi_A,\xi_B)$, we have the relations
\begin{eqnarray}
P(N_A,N_B,t) & \rightarrow & \Pi(\xi_A,\xi_B,t), \\
\Omega^{\frac{1}{2}} \frac{\partial}{\partial\, N_i} P(N_A,N_B,t) &
= & \frac{\partial}{\partial\, \xi_i} \Pi(\xi_A,\xi_B,t).
\end{eqnarray}

Putting everything together, and considering contributions up to
order $\Omega^{\frac{1}{2}}$, yields the following two coupled
differential equations for the macroscopic behavior
\begin{eqnarray}
\frac{d}{dt} \Psi_A(t) = - \alpha_1 \Psi_A +  \Bigl[ \alpha_2 +
\beta_1 \Psi_A  \Bigr]  \Bigl(\rho - \Psi_A - \Psi_B  \Bigr), \label{macro1} \\
\frac{d}{dt} \Psi_B(t) = - \alpha_3 \Psi_B +  \Bigl[ \alpha_4 +
\beta_2 \Psi_B  \Bigr]  \Bigl(\rho - \Psi_A - \Psi_B  \Bigr)
\label{macro2} .
\end{eqnarray}
It can be proved that the last set of equations has a unique
(physically sound) stationary solution, i.e. a unique attractor
\begin{eqnarray}
\Psi_A(t \to \infty) &=& \Psi_A^{st} \nonumber \\
\Psi_B(t \to \infty) &=& \Psi_B^{st} \nonumber.
\end{eqnarray}
This is the main condition to validate the application of van
Kampen's $\Omega$-expansion approach \cite{vKamp}.

The following order, that is $\Omega^{0}$, yields the Fokker-Planck
equation (FPE) governing the fluctuations around the macroscopic
behavior. It is given by
\begin{eqnarray}\label{fpef}
\frac{\partial}{\partial \,t} \Pi(\xi_A,\xi_B,t) = & &
\frac{\partial}{\partial \xi_A} \Bigl[ \left(\alpha_1 \xi_A +
(\alpha_2 + \beta_1 \Psi_A)(\xi_A + \xi_B) - \beta_1 \xi_A (\rho -
\Psi_A - \Psi_B) \right)\Pi(\xi_A,\xi_B,t) \Bigr] \nonumber \\
& + & \frac{\partial}{\partial \xi_B} \Bigl[ \left(\alpha_3 \xi_B +
(\alpha_4 + \beta_2 \Psi_B)(\xi_A + \xi_B) - \beta_2 \xi_B (\rho -
\Psi_A - \Psi_B) \right)\Pi(\xi_A,\xi_B,t) \Bigr] \nonumber \\
& + & \frac{1}{2}\Bigl[ \alpha_1 \Psi_A + (\alpha_2 + \beta_1
\Psi_A)(\rho - \Psi_A - \Psi_B)\Bigr] \frac{\partial^2}{\partial
\xi_A^2}\Pi(\xi_A,\xi_B,t) \nonumber \\
& + & \frac{1}{2}\Bigl[ \alpha_3 \Psi_B + (\alpha_4 + \beta_2
\Psi_B)(\rho - \Psi_A - \Psi_B)\Bigr] \frac{\partial^2}{\partial
\xi_B^2}\Pi(\xi_A,\xi_B,t).
\end{eqnarray}
As is well known for this approach \cite{vKamp}, the solution of
this FPE will have a Gaussian form determined by the first and
second moments of the fluctuations. Hence, in the next section we
analyze the equations governing those quantities.

\section{Behavior of fluctuations}

From the FPE indicated above (Eq. (\ref{fpef})), it is possible to
obtain equations for the mean value of the fluctuations as well as
for the correlations of those fluctuations. For the fluctuations,
$\langle \xi_A (t)\rangle = \eta_A$ and $\langle \xi_B (t)\rangle =
\eta_B$, we have
\begin{eqnarray}
\frac{d}{dt} \eta_A(t) & = & - \Bigl[ \alpha_1 + \alpha_2 + \beta_1
(2 \Psi_A + \Psi_B) - \beta_1 \rho \Bigr] \eta_A - (\alpha_2 +
\beta_1 \Psi_A) \eta_B \\
\frac{d}{dt} \eta_B(t) & = & - \Bigl[ \alpha_3 + \alpha_4 + \beta_2
(\Psi_A + 2 \Psi_B) - \beta_2 \rho \Bigr] \eta_B - (\alpha_4 +
\beta_2 \Psi_B) \eta_A.
\end{eqnarray}

Calling $\sigma_A = \langle \xi_A (t)^2 \rangle$, $\sigma_B =
\langle \xi_B (t)^2 \rangle$, and $\sigma_{AB} = \langle \xi_A
(t)\xi_B (t) \rangle$, we obtain for the correlation of fluctuations
\begin{eqnarray}
\frac{d}{dt} \sigma_A(t) & = & - 2 \alpha_1 \sigma_A - 2 [ \alpha_2
+ \beta_1 \Psi_A ] [\sigma_A + \sigma_{AB}] + 2 \beta_1 \sigma_{A}
[\rho - \Psi_A - \Psi_B] \nonumber \\
& & \,\,\,\,\,\,\,\,\,\,\,\,\,\,\,\, + [ \alpha_1 \Psi_A + (\alpha_2
+ \beta_1 \Psi_A)(\rho - \Psi_A - \Psi_B)], \\
\frac{d}{dt} \sigma_B(t) & = & - 2 \alpha_3 \sigma_B - 2 [\alpha_4 +
\beta_2 \Psi_B ] [\sigma_{AB} + \sigma_B] + 2 \beta_2 \sigma_{B}
[\rho - \Psi_A - \Psi_B] \nonumber \\
& & \,\,\,\,\,\,\,\,\,\,\,\,\,\,\,\, + [ \alpha_3 \Psi_B + (\alpha_4
+ \beta_2 \Psi_B)(\rho - \Psi_A - \Psi_B)], \\
\frac{d}{dt} \sigma_{AB}(t) & = & - [\alpha_1 + \alpha_3]
\sigma_{AB} - [\alpha_2 + \beta_1 \Psi_A ][\sigma_{AB} + \sigma_{B}]
\nonumber \\
& & \,\,\,\,\,\,\,\,\,\,\,\,\,\,\,\, - [ \alpha_4 + \beta_2 \Psi_B ]
[\sigma_A + \sigma_{AB}] + [\rho - \Psi_A - \Psi_B][\beta_1 +
\beta_2] \sigma_{AB}.
\end{eqnarray}

\subsection{Reference state: symmetric case}

Here we particularize the above indicated equations to the
symmetrical case, i.e. the case when $\Psi_A^{st}=\Psi_B^{st}$.
Hence, we adopt
$$\alpha_1 = \alpha_3 = \alpha,\,\,\,\,\, \alpha_2 = \alpha_4 = \alpha',$$
and
$$\beta_1 = \beta_2 = \beta.$$
In such a case, the macroscopic equations (\ref{macro1}) and
(\ref{macro2}) take the form
\begin{eqnarray}
\frac{d}{dt} \Psi_A(t) & = & - [\alpha + \alpha' - \beta] \Psi_A -
\beta \Psi_A^2 - \beta \Psi_A \Psi_B - \alpha' \Psi_B + \alpha'
\label{macro1b} \\
\frac{d}{dt} \Psi_B(t) & = & - [\alpha + \alpha' - \beta] \Psi_B -
\beta \Psi_B^2 - \beta \Psi_A \Psi_B - \alpha' \Psi_A + \alpha'.
\label{macro2b}
\end{eqnarray}
In order to make more explicit the solution of these equations, we
work with the auxiliary variables $\Sigma = \Psi_A + \Psi_B$ and
$\Delta = \Psi_A - \Psi_B$, and use $\rho = 1$. Hence, the last
equations transform now into
\begin{eqnarray}
\frac{d}{dt} \Sigma(t) & = & - \Bigl[\alpha + 2 \alpha' - \beta
\Bigr]
\, \Sigma - \beta \Sigma^2 + 2 \alpha' \label{macro1c} \\
\frac{d}{dt} \Delta(t) & = & - \Bigl[\alpha - \beta \Bigr] \, \Delta
- \beta \Delta \Sigma. \label{macro2c}
\end{eqnarray}
In the long time limit, $t \to \infty$, we found on one hand
$$\Delta ^{st} = 0,$$
implying $\Psi_A^{st} = \Psi_B^{st}$, while on the other hand
$$ 0 = \beta \, \Sigma^{2} + \Bigl[ \alpha + 2 \alpha' - \beta \Bigr]
\, \Sigma - 2 \alpha'.$$ This polynomial has two roots, but only one
is physically sound, namely
\begin{eqnarray}
\Sigma^{st} = \frac{\alpha + 2 \alpha' - \beta}{2 \beta} \left( -1 +
\sqrt{1 + \frac{8 \alpha' \beta}{[\alpha + 2 \alpha' - \beta]^2}}
\right), \label{eqstat1}
\end{eqnarray}
yielding $\Psi_A^{st} = \Psi_B^{st} = \Psi_o^{st} = \frac{1}{2}
\Sigma^{st}.$

In a similar way, we can also simplify the equations for $\eta_A$
and $\eta_B$, calling $S(t) = \eta_A + \eta_B$ and $D(t) = \eta_A -
\eta_B$. The corresponding equations are then rewritten as
\begin{eqnarray}
\frac{d}{dt} S(t) & = & - \Bigl[ \alpha + 2 \alpha' + 2 \beta
(\Psi_A + \Psi_B) - \beta \Bigr]\,S, \label{eqS1}\\
\frac{d}{dt} D(t) & = & - \Bigl[ \alpha + \beta (\Psi_A + \Psi_B) -
\beta \Bigr] D - \beta \Bigl[\Psi_a - \Psi_B \Bigr] S, \label{eqD1},
\end{eqnarray}
while for the correlation of the fluctuations we have
\begin{eqnarray}
\frac{d}{dt} \sigma_A(t) & = & - 2 \alpha \sigma_A - 2 [\alpha' +
\beta \Psi_A ] [\sigma_A + \sigma_{AB}] + 2 \beta [1 - \Psi_A -
\Psi_B]\sigma_{A} \nonumber \\
& & \,\,\,\,\, + \left[ \alpha \Psi_A + (\alpha' + \beta \Psi_A)
(1 - \Psi_A - \Psi_B )\right], \label{sig1}\\
\frac{d}{dt} \sigma_B(t) & = & - 2 \alpha \sigma_B - 2 [\alpha' +
\beta \Psi_B ] [\sigma_{AB} + \sigma_B] + 2 \beta [1 - \Psi_A -
\Psi_B]\sigma_{B}  \nonumber \\
& & \,\,\,\,\,  + \left[ \alpha \Psi_B + (\alpha' + \beta \Psi_B)
(1 - \Psi_A - \Psi_B )\right], \label{sig2}\\
\frac{d}{dt} \sigma_{AB}(t) & = & - 2 \alpha \sigma_{AB} - [\alpha'
+ \beta \Psi_A ][\sigma_{AB} + \sigma_{B}] \nonumber \\
& & \,\,\,\,\, - [ \alpha' + \beta \Psi_B ] [\sigma_{AB} + \sigma_A]
+ 2 \beta [1 - \Psi_A - \Psi_B] \sigma_{AB}.\label{sig3}
\end{eqnarray}

Equations (\ref{eqS1}) and (\ref{eqD1}) show that, in the asymptotic
limit, i.e. for $t \to \infty$, both, $S=0$ and $D=0$, implying that
$\eta_A^{st}=\eta_B^{st}=0$. However, also in the general (non
symmetric) case we expect to find $\eta_A^{st}=\eta_B^{st}=0$. In
addition, from Eqs. (\ref{sig1}), (\ref{sig2}) and (\ref{sig3}), it
is clear that in general we obtain, again for $t \to \infty$, that
$\sigma_{i}^{st} \neq 0$ ($i=A,B,AB$).

As we have seen, in the symmetric case we have $\Psi_A^{st} =
\Psi_B^{st} = \Psi_o^{st}$, hence it is clear that $\sigma_{A}(t)$
and $\sigma_{B}(t)$ behave in a similar way. And in particular
$\sigma_{A}^{st} = \sigma_{B}^{st} = \sigma_{o}^{st}$. In order to
analyze the typical time for return to the stationary situation
under small perturbations, we assume small perturbations of the form
$\sigma_{i}^{st} \approx \sigma_{o}^{st}+ \delta\sigma_{i}(t)$
($i=A,B$) and $\sigma_{AB}^{st} \approx \sigma_{AB,o}^{st}+
\delta\sigma_{i}(t),$ and fix $\Psi_A^{st} = \Psi_B^{st} =
\Psi_o^{st}.$ We find again that both $\delta\sigma_{A}(t)$ and
$\delta\sigma_{B}(t)$ behave in the same way, and this help us to
reduce the number of equations for the decay of correlations. Hence,
we can put $\delta\sigma_{A}(t) = \delta\sigma_{B}(t) =
\delta\sigma_{o}(t).$ The system driving the correlations becomes
\begin{eqnarray}
\frac{d}{dt} \delta\sigma_{o}(t) & = & - 2 \Bigl[\alpha + \alpha' -
\beta + 3\,\beta \Psi_o^{st} \Bigr] \delta\sigma_{o} - 2 \Bigl[
\alpha'+ \beta \Psi_o^{st}\Bigr] \delta\sigma_{AB} \label{sig1-p} \\
\frac{d}{dt} \delta\sigma_{AB}(t) & = & - 2 \Bigl[\alpha + \alpha' -
\beta + 3\,\beta \Psi_o^{st} \Bigr] \delta\sigma_{AB} - 2 \Bigl[
\alpha' + \beta \Psi_o^{st} \Bigr] \delta\sigma_o. \label{sig3-p}
\end{eqnarray}
Clearly, $\delta\sigma_o^{st} = \delta\sigma_{ab}^{st} \equiv 0.$
After some algebraic steps we obtain
\begin{eqnarray}
\delta\sigma_{o}(t) & \simeq & \delta \sigma_{o}(0) \, \exp{\Bigl[
- 2 [\alpha + 2 \beta  \Psi_o^{st} - \beta] \, t  \Bigr]} \label{sig1-p2} \\
\delta\sigma_{AB}(t) & \simeq & \delta \sigma_{AB}(0)\,\exp{\Bigl[
-2 [\alpha + 2 \beta  \Psi_o^{st} - \beta] \, t  \Bigr]}.
\label{sig3-p2}
\end{eqnarray}
These results indicate that, for the symmetrical case, the typical
relaxation time is given by
\begin{eqnarray}
\tau_{relax} = \frac{1}{2} [\alpha + 2 \beta \Psi_o^{st} -
\beta]^{-1}.
\end{eqnarray}

\subsection{Beyond the symmetric case}

Let us call $\alpha_o$, $\alpha_o'$ and $\beta_o$ to the parameter's
values corresponding to the symmetric case. We consider now the
following cases where we vary the parameters
$$\beta_1 = \beta_o, \,\,\,\,\,\, \beta_2 = \beta_o + \Delta\beta,$$
$$\alpha_1 = \alpha_o, \,\,\,\,\,\, \alpha_3 = \alpha_o + \Delta\alpha,$$
$$\alpha_2 = \alpha_o', \,\,\,\,\,\, \alpha_4 = \alpha_o' + \Delta\alpha'.$$
We will vary only one of these parameters, while keeping the rest
fixed. In the following section we present the results (mainly
numerical) corresponding to those different cases.

\section{Results}

As indicated above, the macroscopic equations (Eqs. (\ref{macro1})
and (\ref{macro2})) have a unique attractor, indicating that it is
adequate to apply van Kampen's expansion approach. In this section
we will present some results corresponding to symmetric and
asymmetric situations, that show some typical behavior to be
expected from the model and the approximation method. In what
follows, al parameters are measured in arbitrary units.

\begin{figure}[ht]
\includegraphics[width=15cm,angle=0]{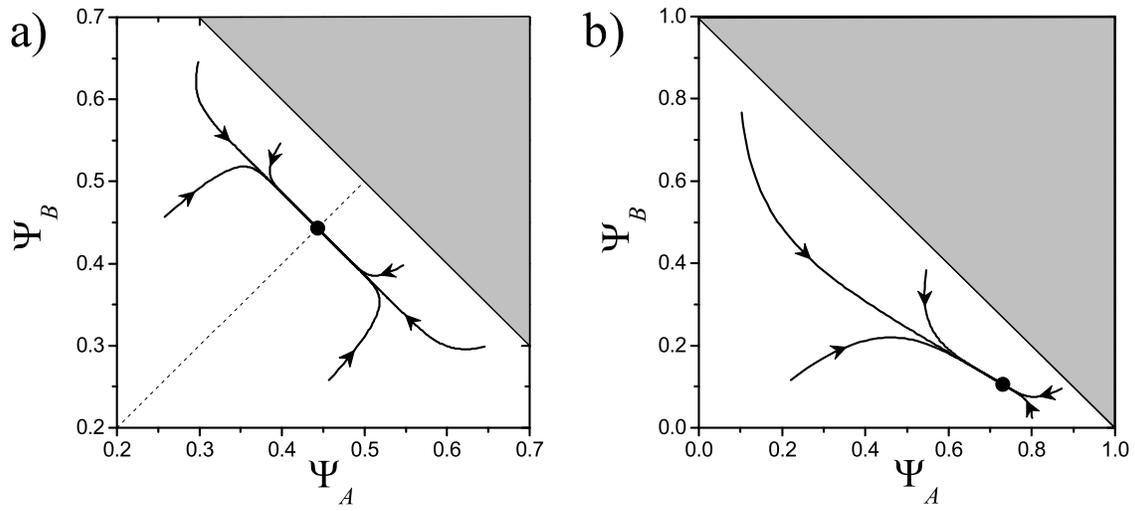}
\caption{Evolution of the macroscopic solutions
(Eqs.(\ref{macro1},\ref{macro2})). Case (a) corresponds to
trajectories towards a symmetric solution (i.e. with $\Psi_A^{st} =
\Psi_B^{st}$), with parameters $\alpha_1 = \alpha_3 = 1$, $\alpha_2
= \alpha_4 = 3$, and $\beta_1 = \beta_2 = 2.$ Case (b) corresponds
to trajectories towards an asymmetric solution (i.e. with
$\Psi_A^{st} \neq \Psi_B^{st}$), with parameters $\alpha_1 = 1$,
$\alpha_3 = 5$, $\alpha_2 = \alpha_4 = 3$, and $\beta_1 = \beta_2 =
2.$ } \label{figmacro}
\end{figure}

In Fig. \ref{figmacro} we show the evolution of $\Psi_A(t)$ and
$\Psi_B(t)$, the macroscopic solutions, indicating some trajectories
towards the attractor: (a) for a symmetric, and (b) an asymmetric
case. It is worth recalling that $\Psi_A$ and $\Psi_B$ are the
density of supporters of party $A$ and party $B$, respectively.
During the evolution towards the attractor, starting from arbitrary
initial conditions, we observe the possibility of a marked initial
increase of the macroscopic density for one of the parties, follow
by a marked reduction, or other situations showing only a decrease
of an initial high density. Such cases indicate the need of taking
with care the results of surveys and polls during, say, an electoral
process. It is possible that an impressive initial increase in the
support of a party can be followed for an also impressive decay of
such a support.

We remark that, due to the symmetry of the problem, it is equivalent
to vary the set of parameters ($\alpha_3,\alpha_4,\beta_2$) or the
set ($\alpha_1,\alpha_2,\beta_1$). Also worth remarking is That in
both panels of Fig \ref{figmacro} the sum of $\Psi_A$ and $\Psi_B$
is always $<1$, so verifying that there is always a finite fraction
of undecided agents.

\begin{figure}[ht]
\includegraphics[width=17cm,angle=0]{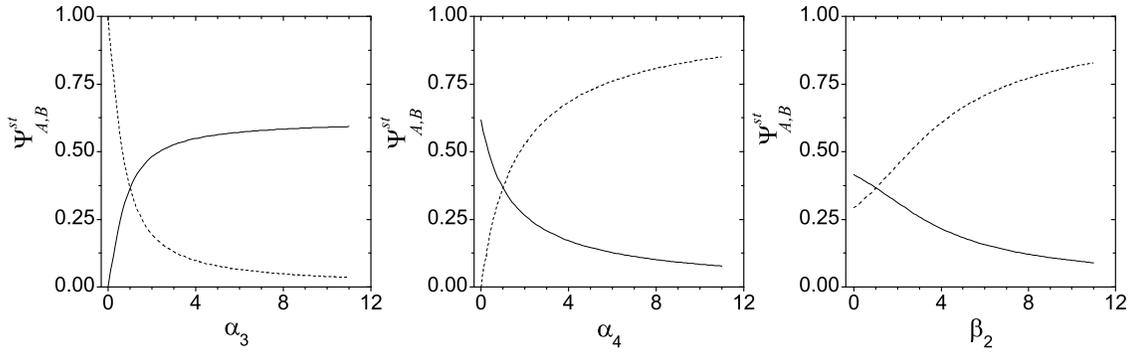}
\caption{Dependence of the stationary macroscopic solutions on
different system parameters: (a) on $\alpha_3$, the rest of
parameters are $\alpha_1 = \alpha_2 = \alpha_4 = 1$, and $\beta_1 =
\beta_2 = 1.$ (b) on $\alpha_4$, the rest of parameters are
$\alpha_1 = \alpha_2 = \alpha_3 = 1$, and $\beta_1 = \beta_2 = 1.$
(c) on $\beta_2$, the rest of parameters are $\alpha_1 = \alpha_2 =
\alpha_3 = \alpha_4 = 1$, and $\beta_1 = 1.$ In all three cases, the
continuous line corresponds to $\Psi_A^{st}$ while $\Psi_B^{st}$ is
indicated by the dotted line.} \label{figmacro2-1}
\end{figure}

In Fig. \ref{figmacro2-1} we depict the dependence of the stationary
macroscopic solutions on different parameters of the system. On Fig.
\ref{figmacro2-1}(a) the dependence on $\alpha_3$ is represented. It
is apparent that for $\alpha_3<\alpha_1$, we have
$\Psi_B^{st}<\Psi_A^{st}$, while for $\alpha_3>\alpha_1$, we find
the inverse situation. Clearly, $\Psi_B^{st}=\Psi_A^{st}$ when
$\alpha_3 = 1 (=\alpha_1)$, as it corresponds to the symmetric case.
Similarly, in Figs. \ref{figmacro2-1}(b) and \ref{figmacro2-1}(c) we
see the dependence of the stationary macroscopic solutions on the
parameters $\alpha_4$ and $\beta_2$, respectively. Also in these
cases we observe similar behavior as in the previous one, when
varying the indicated parameters. The parameters $\alpha_3$ or
$\alpha_4$ (and similarly for $\alpha_1$ or $\alpha_2$) correspond
to spontaneous changes of opinion, and may be related to a kind of
\textit{social temperature} \cite{babinec,weidlich2,last}. However,
also $\beta_1$ and $\beta_2$ are affected by such a temperature. So,
the variation of these parameters in Fig. \ref{figmacro2-1}
correspond to changes in the social temperature, changes that could
be attributed, in a period of time preceding an election, to
increase in the level of discussions as well as the amount of
propaganda.

\begin{figure}[ht]
\includegraphics[width=17cm,angle=0]{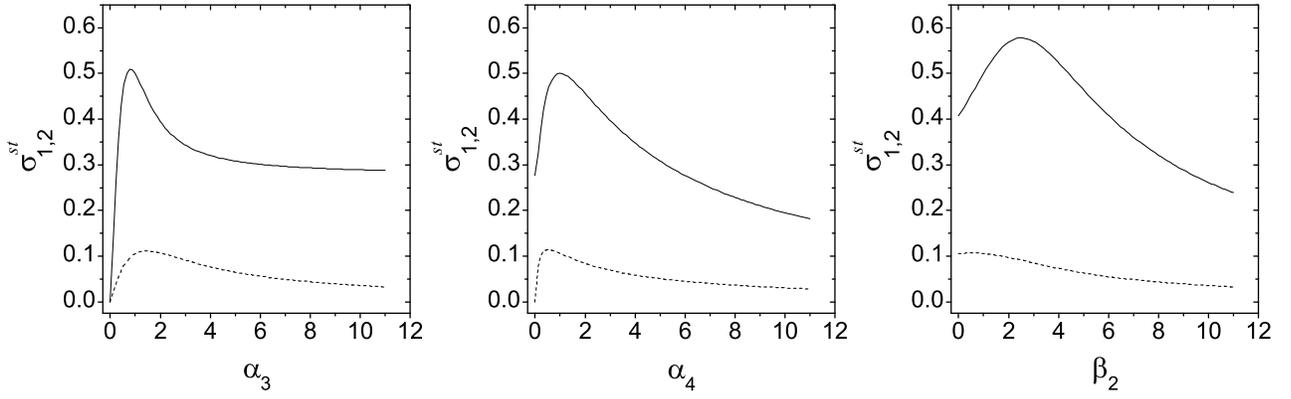}
\caption{Dependence of the stationary correlation functions
$\sigma_{i}$ (with $i=1,2$) corresponding to the projection of
$\sigma_{A,B,AB}$ on the principal axes, on different parameters of
the system: (a) on $\alpha_3$, the other parameters are $\alpha_1 =
\alpha_2 = \alpha_3 = 1$, and $\beta_1 = \beta_2 = 1.$ (b) on
$\alpha_4$, the other parameters are $\alpha_1 = \alpha_2 = \alpha_3
= 1$, and $\beta_1 = \beta_2 = 1.$ (b) on $\beta_2$, the other
parameters are $\alpha_1 = \alpha_2 = \alpha_3 = \alpha_4 = 1$, and
$\beta_1 = 1.$} \label{figsigb-3}
\end{figure}

In Fig. \ref{figsigb-3} we depict the dependence of the stationary
correlation functions for the fluctuations $\sigma_{i}$ (with
$i=1,2$, corresponding to the projection of $\sigma_{A,B,AB}$ on the
principal axes), on different systems' parameters. In Fig.
\ref{figsigb-3}(a) the dependence on $\alpha_3$ is represented, and
similarly in Figs. \ref{figsigb-3}(b) and \ref{figsigb-3}(c), the
dependence on the parameters $\alpha_4$ and $\beta_2$, respectively.
We observe that, as the parameters are varied (that, in the case of
$\alpha_3$ and $\alpha_4$, and as indicated above, could be
associated to a variation of the \textit{social temperature}) a
\textbf{tendency inversion} could arise. This indicates that the
dispersion of the probability distribution could change with a
variation of the \textit{ social temperature}. This is again a
warning for taking with some care the results of surveys and polls
previous to an electoral process.

\begin{figure}[ht]
\includegraphics[width=17cm,angle=0]{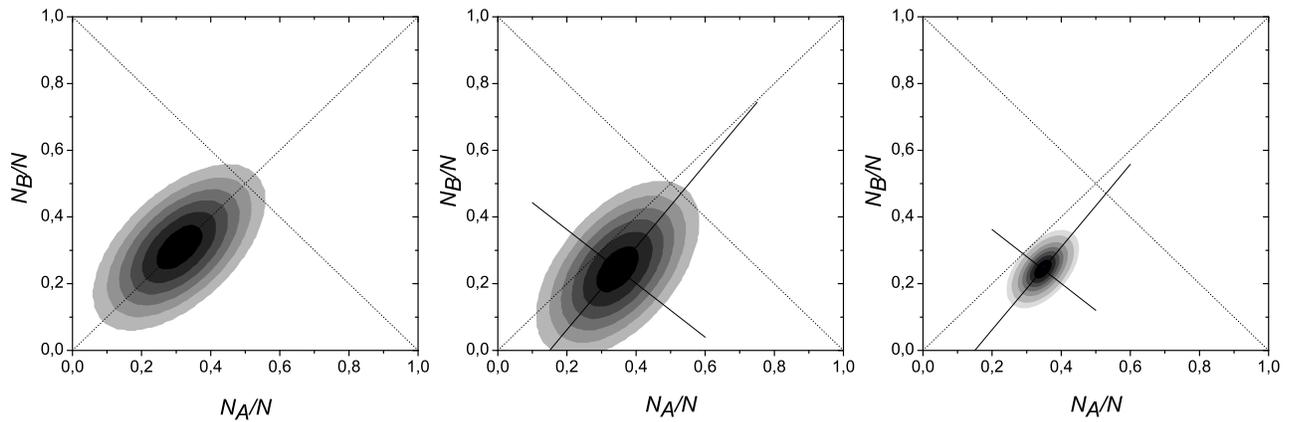}
\caption{Stationary, Gaussian, probability distribution
$\Pi(\xi_A,\xi_B)^{st}$ projected on the original $(N_A, N_B)$
plane. On the let we have a symmetrical case with $\alpha_1 =
\alpha_3 = 2$, $\alpha_2=\alpha_4=1$, $\beta_1 = \beta_2 = 2$, and
the population is $N = 100$. The central plot shows an asymmetrical
case, with $\alpha_1=2$ and $\alpha_3=2.5$, while
$\alpha_2=\alpha_4=1$, $\beta_1=\beta_2=2$, and the population is
$N=100$. On the right we have the same asymmetrical case as before,
but now $N=1000$, showing the dispersion's reduction of the Gaussian
distribution.} \label{figpdf-1}
\end{figure}

Figure \ref{figpdf-1} shows the stationary (Gaussian) probability
distribution (pdf) $\Pi(\xi_A,\xi_B)^{st}$ projected on the original
$(N_A, N_B)$ plane. We show three cases: on the left a symmetrical
case, the central one corresponds to an asymmetrical situation with
a population of $N=100$, and on the right the same asymmetrical
situation but with a population of $N=1000$. This last case clearly
shows the influence of the population number in reducing the
dispersion (as the population increases). We can use this pdf in
order to estimate the probability $p_{i}$ $(i=A,B)$, of winning for
one or the other party. It corresponds to the volume of the
distribution remaining above, or below, the bisectrix $N_A/N =
N_B/N.$  In the symmetrical case, as is obvious, we obtain
$p_{A}=p_{B}=0.5$ (or $50\%$), while in the asymmetrical case we
found $p_{B}=0.257$ (or $25.7\%$) and $p_{B}=0.015$ (or $1.5\%$) for
$N=100$ and $N=1000$, respectively. These results indicate that, for
an asymmetrical situation like the one indicated here, we have a non
zero probability that the minority party could, due to a fluctuation
during the voting day, win the election. However, in agreement with
intuition, as far as $N \gg 1$, and the stationary macroscopic
solution departs from the symmetric case, such a probability $p_{i}$
reduces proportionally to $N^{-1}$ \cite{comment}.

\begin{figure}[ht]
\includegraphics[width=17cm,angle=0]{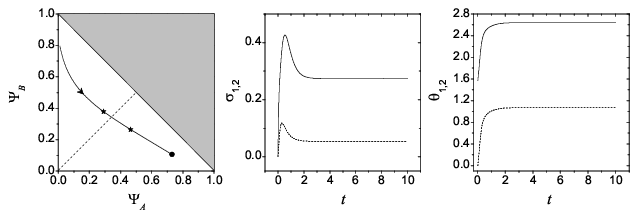}
\caption{On the left, we have the time evolution of the macroscopic
solutions $\Psi_A(t)$ and $\Psi_B(t)$. The parameter values are
$\alpha_1 = 1$, $\alpha_3 = 5$, $\alpha_2=\alpha_4 = 3$, $\beta_1 =
\beta_2 = 2$. The stars indicate the position where the maxima that
appear in the next panel occurs. Central part, time evolution of the
correlation functions $\sigma_{i}$ (with $i=1,2$) corresponding to
the projection of $\sigma_{A,B,AB}$ on the principal axes. On the
right, the angle between the principal axes and the figure axes. The
parameters are $\alpha_1 = 1$, $\alpha_3 = 5$, $\alpha_2 = \alpha_4
= 3$, and $\beta_1 = \beta_2 = 2.$} \label{figsig-4}
\end{figure}

In Fig. \ref{figsig-4}, on the left,  we show a typical result for
the time evolution of the macroscopic solution towards an asymmetric
stationary case. In the same figure, in the central part we find the
associated time evolution of the correlation functions for the
fluctuations, $\sigma_{i}$ (with $i=1,2$) corresponding to the
projection of $\sigma_{A,B,AB}$ on the principal axes, while on the
right we show the evolution of the angle between the principal axes
and the figure axes. The temporal reentrance effect that has been
observed in other studies exploiting the van Kampen's approach
\cite{vKamp,ich-1} is apparent. This is a new warning, indicating
the need to take with some care the results of surveys and polls
during an electoral process.

\begin{figure}[ht]
\includegraphics[width=16cm,angle=0]{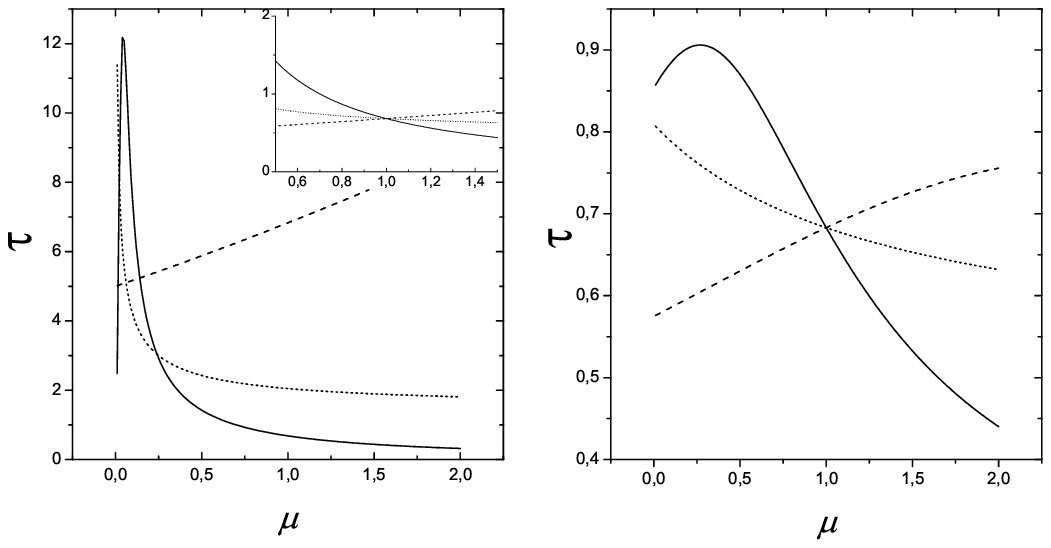}
\caption{Dependence of the \textit{dominant} relaxation time on
different system parameters. On the left, symmetrical case:
continuous line varying $\alpha_1=\alpha_3$, dotted line varying
$\alpha_2=\alpha_4$, and dashed line varying $\beta_1=\beta_2$. In
order to compare all three, the dotted line was multiplied by 3,
while the dashed one by 10. The inset shows, now on the same scale,
the crossing of the lines at the point where all the parameters are
equal to 1. On the right, asymmetrical case: continuous line varying
$\alpha_1$, dotted line varying $\alpha_2$, and dashed line varying
$\beta_1$. In all cases, the parameters that remain constant are all
$=1$.} \label{relaxation-1}
\end{figure}

In Fig. \ref{relaxation-1} we depict the dependence of the dominant
(or relevant) relaxation time, that is the slowest of the three
relaxation times, on different parameters of the system. On the
left, we show a symmetrical case where the different lines represent
the dependence respect to variation of: $\alpha_1=\alpha_3$
indicated by a continuous line; $\alpha_2=\alpha_4$ indicated by
dotted line; $\beta_1=\beta_2$ indicated by dashed line. The strong
dependence of the relaxation time on $\alpha=\alpha_1=\alpha_3$ is
apparent (in order to be represented in the same scale, the other
two cases are multiplied by 3 or 10, respectively). This means that
changes in the \textit{social temperature} that, as discussed
before, induce changes in $\alpha (=\alpha_1=\alpha_3)$, could
significatively change the dominant relaxation time. On the right we
show an asymmetrical case where, as before, the different lines
represent the dependence respect to variation of: $\alpha_1$,
indicated by a continuous line; $\alpha_2$, indicated by a dotted
line; and $\beta_1$, indicated by dashed line. It is worth remarking
that, when all the the parameters ($\alpha_1$, $\alpha_2$ and
$\beta_1$) are equal to 1, we see that the relaxation time is the
same. On the left figure, this is shown in the inset. In the
asymmetrical case, the behavior is of the same order for the
variation of the three parameters. However, the comment about the
effect of changes in the \textit{social temperature} remain valid.

\section{Conclusions}

We have studied a simple opinion formation model (that is a
\textit{toy model}), analogous to the one studied in \cite{redner3}.
It consists of two parties, $A$ and $B$, and an intermediate group
$I$, that we call \textit{undecided agents}. It was assumed that the
supporters of parties $A$ and $B$ do not interact among them, but
only through their interaction with the group $I$, convincing its
members through a mean-field treatment; that members of $I$ are not
able to convince those of $A$ or $B$, but instead we consider a
nonzero probability of a spontaneous change of opinion from $I$ to
the other two parties and viceversa. It is this possibility of
spontaneous change of opinion that inhibits the possibility of
reaching a consensus, and yields that each party has some
statistical density of supporters, as well as a statistical
stationary number of undecided agents.

Starting from the master equation for this toy model, the van
Kampen's $\Omega$-expansion approach \cite{vKamp} was exploited in
order to obtain the \textit{macroscopic} evolution equations for the
density of supporters of $A$ and $B$ parties, as well as the
Fokker-Planck equation governing the fluctuations around such a
macroscopic behavior. Through this same approach information about
the typical relaxation behavior of small perturbations around the
stationary macroscopic solutions was obtained.

The results indicate that one needs to take with care the results of
social surveys and polls in the months preceding an electoral
process. As we have found, it is possible that an impressive initial
increase in the support of a party can be followed for an also
impressive decay of such a support. The dependence of the
macroscopic solutions as well as the correlation of the fluctuations
on the model parameters, variation in $\alpha_3$, $\alpha_4$ or $
\beta_2$ (that, due to the symmetry of the model are similar to
varying $\alpha_1$, $\alpha_2$ or $\beta_1$) was also analyzed. As
the parameters $\alpha_i$ correspond to spontaneous change of
opinion, or $ \beta_i$ to convincing capacity, and it is possible to
assume that have an ``activation-like structure", we can argue that
this could be related to changes in the \textit{social temperature},
and that such a temperature could be varied, for instance, in a
period near elections when the level of discussion as well as the
amount of propaganda increases.

We have also analyzed the probability that, due to a fluctuation,
the minority party could win a loose election, and that such a
probability behaves inversely to $N$ (the population number). Also
analyzing the temporal behavior of the fluctuations some ``tendency
inversion" indicating that, an initial increase of the dispersion
could be reduced as time elapses was found.

We have also analyzed the relaxation of small perturbations near the
stationary state, and the dependence of the typical relaxation times
on the system parameters was obtained. This could shead some light
on the social response to small perturbations like an increase of
propaganda, or dissemination of information about some ``negative"
aspects of a candidate, etc. However, such an analysis is only valid
near the macroscopic stationary state, but looses its validity for a
very large perturbation. For instance, a situation like the one
lived in Spain during the last elections (the terrorist attack in
Madrid on March 11, 2003, just four days before the election day),
clearly was a very large perturbation that cannot be described by
this simplified approach.

Finally, it is worth to comment on the effect of including a direct
interaction between both parties $A$ and $B$. As long as the direct
interaction parameter remains small, the monostability will persist,
and the analysis, with small variations will remain valid. However,
as the interaction parameter overcomes some threshold value, a
transition towards a bistability situation arise, invalidating the
exploitation of the van Kampen's $\Omega$-expansion approach.

\begin{acknowledgments}
We acknowledge financial support from Ministerio de Educaci\'on y
Ciencia (Spain) through Grant No. BFM2003-07749-C05-03 (Spain). MSL
and IGS are supported by a FPU and a FPI fellowships respectively
(Spain). JRI acknowledges support from FAPERGS and CNPq, Brazil, and
the kind hospitality of Instituto de F\'{\i}sica de Cantabria and
Departamento CITIMAC, Universidad de Cantabria, during the initial
stages of this work. HSW thanks to the European Commission for the
award of a {\it Marie Curie Chair} at the Universidad de Cantabria,
Spain.

\end{acknowledgments}


\begin{thebibliography}{99}
\bibitem{comment} It is worth commenting that it is convenient to
avoid those patological range of parameters making that $\Upsilon =
\Psi_A^{st} + \Psi_B^{st}$ falls within a very thin strip near the
frontiers of the physical region (i.e. the region limited by
$\Upsilon = 1$, or $\Psi_A^{st} = 0$, or $\Psi_B^{st} = 0$). In such
cases, the tail of fluctuations falling outside the physical region
will be too large invalidating the whole approach. Clearly, the
parameters choosen for Fig. \ref{figpdf-1} avoid such patological
situation, as the fluctuation tails falling outside the physical
region are negligible.
\end{thebibliography}
\end{document}